\documentclass[journal]{IEEEtran}
\usepackage{amsmath,amssymb}
\usepackage{graphicx}
\usepackage{epstopdf}
\usepackage{color}
\usepackage{subcaption}
\usepackage{stfloats}
\usepackage{caption}
\usepackage{array}
\usepackage{enumerate}
\usepackage{makecell}
\usepackage{colortbl}
\usepackage{hyperref}
\usepackage{cleveref}
\usepackage[stable]{footmisc}

\usepackage{fancyhdr}
\usepackage{tikz}
\newcommand\copyrighttext{%
  \footnotesize \textcopyright 2017 IEEE. Personal use of this material is permitted.
  Permission from IEEE must be obtained for all other uses, in any current or future
  media, including reprinting/republishing this material for advertising or promotional
  purposes, creating new collective works, for resale or redistribution to servers or
  lists, or reuse of any copyrighted component of this work in other works.
  DOI: \href{https://doi.org/10.1109/TIM.2017.2745081}{10.1109/TIM.2017.2745081}}
\newcommand\copyrightnotice{%
\begin{tikzpicture}[remember picture,overlay]
\node[anchor=south,yshift=0pt] at (current page.south) {\fbox{\parbox{\dimexpr\textwidth-\fboxsep-\fboxrule\relax}{\copyrighttext}}};
\end{tikzpicture}%
}

\usepackage[mathlines,switch]{lineno}

\begin{document}
		\title{Nonlinear Cuff-less Blood Pressure Estimation of Healthy Subjects Using Pulse Transit Time and Arrival Time
		}
		
	\author{Amirhossein~Esmaili,~\IEEEmembership{Student Member,~IEEE,} Mohammad~Kachuee,~\IEEEmembership{Student Member,~IEEE,} and ~Mahdi~Shabany,~\IEEEmembership{Member,~IEEE}
	\thanks{\textbf{The collected data set can be accessed using the following url link: \url{www.kaggle.com/mkachuee/noninvasivebp}.} The authors are with the Department of Electrical Engineering, Sharif University of Technology, Tehran 1458889694, Iran (e-mail: amirhossein.ed12@gmail.com; m.kachuee@gmail.com; mahdi@sharif.edu).}
	}
\maketitle
\copyrightnotice

\textcolor[rgb]{0.00,0.00,0.00}{
\begin{abstract}
This paper presents a novel blood pressure (BP) estimation method based on pulse transit time (PTT) and pulse arrival time (PAT) to estimate the systolic blood pressure (SBP) and diastolic blood pressure (DBP). A data acquisition hardware is designed for high-resolution sampling of phonocardiogram (PCG), photoplethysmogram (PPG) and electrocardiogram (ECG). PCG and ECG perform as the proximal timing reference to obtain PTT and PAT indexes, respectively. In order to derive a BP estimator model, a calibration procedure including a supervised physical exercise is conducted for each individual which causes changes in their BP and then, a number of reference BPs are measured alongside the acquisition of the signals per subject. It is suggested to use a force-sensing resistor (FSR) that is placed under the cuff of the BP reference device to mark the exact moments of reference BP measurements, which are corresponding to the inflation of the cuff. Additionally, a novel BP estimator nonlinear model, based on the theory of elastic tubes, is introduced to estimate the BP using PTT/PAT values precisely. \textcolor{black}{The proposed method is evaluated on 32 subjects. Using the PTT index, the correlation coefficients for SBP and DBP estimation are 0.89 and 0.84, respectively. Using the PAT index, the correlation coefficients for SBP and DBP estimation are 0.95 and 0.84, respectively.} The results show that the proposed method, exploiting the introduced nonlinear model with the use of PAT index or PTT index, provides a reliable estimation of SBP and DBP.\\
\end{abstract}
}


\section{Introduction} \label{section.introduction}
\textcolor[rgb]{0.00,0.00,0.00}{Hypertension, or long-term high blood pressure (BP), is a primary factor increasing the risk of ischemic heart disease, heart failure, aortic aneurysms and so forth \cite{prospective2002age}, \cite{lip2007abc}.}
\textcolor{black}{hypertension is prevalent among about 26\% and 23\% percent of men and women, respectively. Also, it is more frequent in countries with low GDP values \cite{world2015world}}. These facts necessitate the usage of effective methods for continuous BP monitoring.

\textcolor{black}{BP is defined as the pressure that the blood applies to the vessel walls during the blood circulation. BP is a periodic vital signal with the same period as of heart rate period. Systolic blood pressure (SBP) and diastolic blood pressure (DBP) are defined as the maximum and minimum values of the BP during each period, respectively.}

Mainly, there are two categories for BP measurement, i.e., invasive and non-invasive methods. Although invasive methods are known for providing accurate and
beat-to-beat record of BP, its usage is limited because of the risks of invasiveness that might cause infection and bleeding, so it is generally used in hospitals and intensive care units for hospitalized patients.


\textcolor{black}{Among non-invasive BP measurement methods, auscultatory \cite{perloff1993human} and oscillometry \cite{alpert2014oscillometric} measurement methods are the most prevalent ones.} Nevertheless, they cannot provide beat-to-beat
monitoring of BP since the inflatable cuff, used normally in these approaches, requires inflation and deflation
for each BP measurement, which is time consuming (e.g. about one minute). Moreover, the cuff causes inconvenience for patients due to the occlusion of blood flow in the arteries. 
\textcolor{black}{Also, various factors (mental or physical) may lead to BP changes over time. Therefore, It would be highly desirable to monitor BP continuously and in a non-invasive and cuff-less manner that is convenient.}

\textcolor{black}{A wide variety of research has been conducted in devising methods for the continuous and cuff-less measurement of BP. Among these, using cardiovascular parameters such as pulse wave velocity (PWV) are the most studied ones. PWV is defined as the velocity of the pressure wave that is propagating through arteries. PWV is usually estimated by measuring the PTT. PTT is the time that takes for the pressure wave to travel from a proximal point to a distal point in the arterial tree, within the same cardiac cycle.}
The fundamental principle is that arterial elasticity increases with BP and it also influences the PWV. As a result, BP can be estimated by the measurement of PTT or PWV. The exact relation between BP and PTT is, however, individual-specific and depends on the physical properties of the vessels and blood of each person. \textcolor{black}{Therefore, to approximate this relation, a calibration procedure in which a number of reference BP measurements are done, is required for each individual \cite{kim2015ballistocardiogram}.}

\textcolor{black}{Usually, in the literature, electrocardiogram (ECG) is used as a proximal timing reference and the photoplethysmogram (PPG) is used as a distal timing reference for PTT measurement \cite{gesche2012continuous}, \cite{cattivelli2009noninvasive}, \cite{cheol2013using}, \cite{he2014secondary} \cite{kachuee2015cuff}. However, using ECG and PPG signals leads to the measurement of another cardiovascular parameter that is pulse arrival time (PAT). PAT includes not only the desired PTT but also a pre-ejection period (PEP).} The PAT has shown a high correlation with BP (especially SBP) \cite{wong2009evaluation}. Some papers have used other vital signals, instead of ECG, such as phonocardiogram (PCG) in \cite{chandrasekaran2013cuffless} and \cite{shukla2015noninvasive}, or ballistocardiogram (BCG) in \cite{kim2015ballistocardiogram}, as the proximal timing reference to obtain PTT index instead of PAT index. 

A major portion of research works have used estimator models that assume a simple linear relation between PTT/PAT and BP \cite{cattivelli2009noninvasive}, \cite{chandrasekaran2013cuffless}. However the relationship between PTT/PAT and BP is actually nonlinear and parameters defining this nonlinear relation are individual-specific and depend on physical properties of vessels and blood (see Section \ref{section.background_BP_PTT} for more detalis). The authors in \cite{gesche2012continuous} have assumed this nonlinearity between PAT and BP. However, parameters of this nonlinear relation are determined globally, using a train set of subjects, and the same parameters are employed to estimate BP for a test set of subjects with different physical properties.

In this paper, a novel nonlinear model, based on modeling vessels as elastic tubes, is introduced for BP estimation based on PTT/PAT values. A calibration procedure for each subject, which uses a novel way to mark the moments of reference BP measurements, is utilized to determine the subject-specific parameters of the proposed model per individual. The PCG and ECG signals are employed as the proximal timing references for obtaining PTT index and PAT index, respectively. PPG is used as the distal timing reference for obtaining both PAT and PTT indexes.



The rest of the paper is structured as follows: Section \ref{section.background} explains the physical background behind the relation between PTT and BP in each individual. Section \ref{section.datacollection} describes the sampling hardware setup and the data acquisition procedure employed for each subject. Section \ref{section.processingpipeline} presents the processing pipeline exploited to measure PTT/PAT values from acquired signals. Section \ref{section.results} provides results and compares them with existing related works. \textcolor{black}{Section \ref{section.discussion} presents some discussions on the results and some limitations of the paper.} Finally, Section \ref{section.conclusion} concludes the paper.

\section{Background} \label{section.background}

\subsection{The Relationship between BP and PTT} \label{section.background_BP_PTT}
\textcolor{black}{The key principle} behind using PWV for BP estimation is that the blood flow in the arteries can be modeled as the propagation of pressure waves inside elastic tubes.
Specifically, by assuming vessels like elastic tubes, Elastic modulus ($E$) of the vessel walls can be written as:
\begin{equation}\label{eq.young}
E=E_0 e^{\alpha (P-P_0)},
\end{equation}
where $P$ is the fluid pressure (here BP), and $E_0$, $P_0$ and $\alpha$ are individual-specific parameters \cite{hughes1978measurements}. The compliance ($C$) is defined as the rate at which the tube cross section changes in terms of $P$. Considering the conservation of mass and momentum equations, it can be seen that $C$ is a function of $P$:
\begin{equation}\label{eq:CP}
C(P) = \frac{A_{m}}{\pi P_{1} [1+(\frac{P-P_{0}}{P_{1}})^{2}]} ,
\end{equation}
where $P_{0}$, $P_{1}$, and $A_{m}$ are individual-specific parameters \cite{kachuee2016cuff}. Writing the wave propagation equations inside the vessels, which are assumed as elastic tubes, leads to the following equation \cite{remoissenet2013waves}:
\begin{equation}\label{eq:Pxt}
P(x,t) = f( x \pm t/\sqrt{LC(P)} ) ,
\end{equation}
where $L = \rho/A$; in which, $\rho$ and $A$ are the blood density and vessel cross section, respectively. Therefore, PTT for a tube of length $l$ can be derived as:
\begin{equation}\label{eq:PWV}
PTT =\frac{l}{PWV} =l \sqrt{L C(P)} .
\end{equation}
Furthermore, by substituting $L$ and $C(P)$, PTT can also be formulated as:
\begin{equation} \label{eq.PWV_2}
PTT = l \sqrt{ \frac{\rho A_{m}}{\pi A P_{1} [1+(\frac{P-P_{0}}{P_{1}})^{2}]} } .
\end{equation}
The PTT-BP relationship for each individual is described in (\ref{eq.PWV_2}) with the aid of individual-specific parameters. $l$ is related to the distance between the distal and proximal points at the time of PTT measurement. By solving (\ref{eq.PWV_2}) for $P$ ($BP$ here), $BP$ can be derived as a function of $PTT$:
\begin{equation}\label{eq:math}
BP = P_{0} + \sqrt{-P_{1}^{2} + \frac{l^2\rho A_{m}}{\pi A} \times \frac{1}{PTT^{2}}},
\end{equation}
which can be rewritten in a more simplified form as:
\begin{equation}\label{eq:P}
BP = a_{0} + \sqrt{a_{1} + a_{2} \frac{1}{PTT^{2}}},
\end{equation}
\textcolor{black}{where $a_{0}$, $a_{1}$, and $a_{2}$ are derived for each subject. In other words, instead of directly determining the individual-specific parameters of (\ref{eq:math}), the BP-PTT relationship can be obtained by determining $a_{0}$, $a_{1}$, and $a_{2}$ parameters in (\ref{eq:P}).}

\textcolor[rgb]{0.00,0.00,0.00}{
PAT can also be used instead of PTT for the purpose of cuff-less BP estimation. PAT is defined as PTT plus the PEP time interval, where PEP is the time it takes for the myocardium to raise enough pressure to open the aortic valve and start forcing the blood out of the ventricle. Yet, it has been shown that PAT has a high correlation with BP.
}
%


\textcolor[rgb]{0.00,0.00,0.00}{
\textcolor[rgb]{0.00,0.00,0.00}{In this paper, we have proposed using the non-linear BP-PTT relation of (\ref{eq:P}) to present a more accurate estimation of individual's BP.} 
To derive the constant parameters of \eqref{eq:P} for each individual, a calibration procedure is required in which the subject's BP is varied by means of some interventions \cite{choi2013noninvasive}, \cite{cheol2013using}, \cite{sola2013noninvasive}, \cite{chen2000continuous}. After the interventions, a number of reference BPs are measured while signals for each individual are acquired (see Section \ref{section.datacollection}). Using acquired signals, PTT/PAT values, corresponding to the reference BP measurements are obtained as it is indicated in Section \ref{section.PTT_PAT}. The obtained pairs of PTT/PAT values and reference BP measurements enables us to derive the parameters of \eqref{eq:P} as it is presented in Section \ref{section.model}.
}

\subsection{PTT/PAT Measurement Method}
\begin{figure}[!t]
\centering
\includegraphics[width=0.5\textwidth]{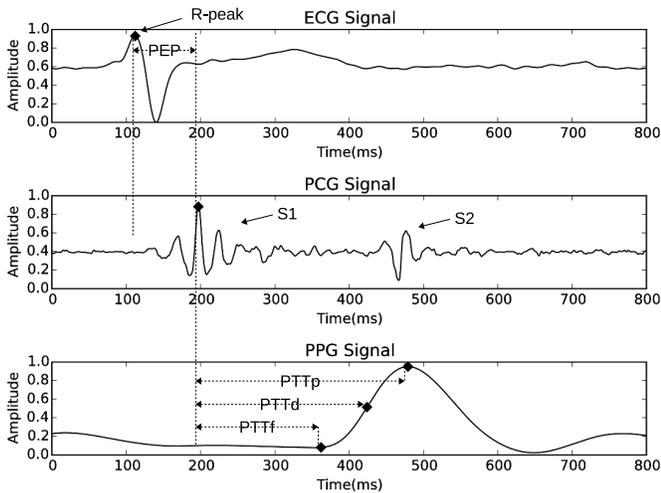}
\caption{{\footnotesize \textcolor[rgb]{0.00,0.00,0.00}{Obtaining different variations of PTT values using PCG S1-peak and PPG different characteristic points. Corresponding PAT values are obtained with the use of ECG R-peak instead of PCG S1-peak as the proximal timing reference, where they are the same as PTT values plus a PEP offset.}}} \label{fig.pttvar}
\end{figure}

Corresponding PTT/PAT to each pair of proximal and distal time instances is measured by the calculation of distal time minus proximal time.
\textcolor[rgb]{0.00,0.00,0.00}{
In this paper, PCG S1-peak is considered as the proximal timing reference to measure PTT, as it approximately represents the moment that blood pressure wave leaves the heart. \textcolor{black}{S1 and S2 are two dominant types of sounds in the PCG signal, where S1 corresponds to the closure of mitral and tricuspid valves, while S2 corresponds to the closure of aortic and pulmonary valves}. Also, similar to many works in literature, ECG R-peak is considered as the proximal timing reference for measuring PAT. (see Fig. \ref{fig.pttvar}).}

\textcolor[rgb]{0.00,0.00,0.00}{
A characteristic point on the PPG, in the same cardiac cycle as the selected proximal timing reference, is considered as the distal timing reference. This point could be either the point where PPG \textcolor{black}{begins} to rise in each cardiac cycle ($PPG_f$), the point where the PPG slope reaches its maximum ($PPG_d$), or the systolic peak of PPG ($PPG_p$), where \textcolor{black}{the PPG itself} reaches its maximum. The corresponding PAT/PTT values are called $PAT_f$, $PAT_d$, $PAT_p$ and $PTT_f$, $PTT_d$, $PTT_p$, respectively (see Fig. \ref{fig.pttvar}).
}

\section{Data Collection Methodology} \label{section.datacollection}
\subsection{Hardware Setup} \label{section.equipment}
\begin{figure}[!t]
	\centering
	\includegraphics[width=0.5\textwidth]{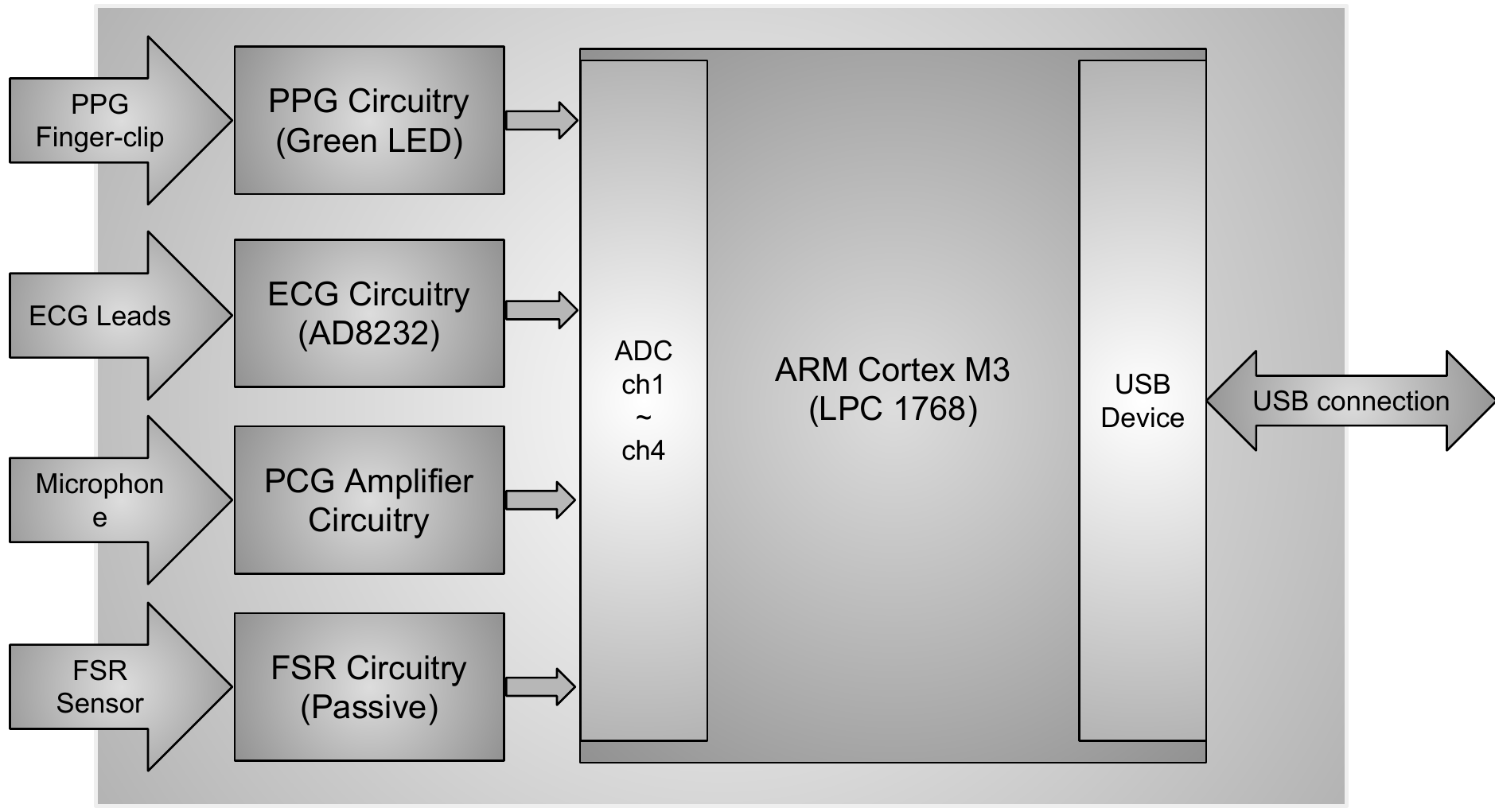}
	\caption{{\footnotesize Block diagram of \textcolor{black}{the data acquisition board.}}} \label{fig.hw}
\end{figure}

\textcolor{black}{The utilized hardware setup in this work includes a data acquisition board developed in our team, a portable computer, and a commercial BP monitor. The board acquires the vital signals (i.e. PCG, PPG and ECG) in addition to the \textcolor[rgb]{0.00,0.00,0.00}{signal from a force-sensing resistor (FSR).} The signals are recorded synchronously from four separate ADC channels using 1KHz sampling frequency. Fig.~\ref{fig.hw} presents the block diagram of the designed data acquisition board. The board and the portable computer are connected via Universal Serial Bus (USB) protocol and the board sends data frames of \mbox{1-second} length to the computer. Accordingly, there is a maximum \mbox{1-second} latency between the channels and the computer, which considerably facilitates appropriate placement of the sensors. \textcolor{black}{In order to ensure that data samples from different channels are acquired and transferred to the computer synchronously, ADC inputs from all channels were read in the burst mode and placed in the data frame in an interleaved fashion. Via the proposed synchronization method, the offset among the four signals at the start of the recording can be assumed to be constant for each specific individual and hardware pair and therefore, the effect of these constant offset values is compensated and cancelled by the calibration procedure.} Also, a software is designed for the computer, which saves the time series data of ADC channels sent from the board as well as storing each subject’s physiological information (i.e.  height, weight, arm length, age, etc.) and reference BP measurements.}



In this setup, reflection PPG with the wavelength of 530 nm (green) is acquired from the finger of the left hand. \textcolor{black}{Reflectance PPG is prefered over its transmissive counterpart due to its convenience of sensor placement for end users.} The AD8232 (fully integrated single-lead ECG front-end) chip was used to acquire high-resolution ECG from the left and right hands (leg drive was not connected). An electret condenser microphone, which is connected to a diaphragm by means of a hollow tube, was used to acquire PCG signal. A chest-belt was used to fix the diaphragm on the subject's chest. An automatic digital BP monitor device (Model M3, OMRON, Japan) was used to measure reference BP values. According to the manufacturer's datasheet, the accuracy of this device is about 3 mmHg, which is compliant with the conventional health standards such as AAMI and BHS. For instance, the mean absolute error of about 5 mmHg is considered to be the acceptable error rate according to the AAMI guidelines \cite{standard_aami}. \textcolor{black}{In addition, before starting the data collection process, we also measured the BP of 7 subjects randomly using both Omron M3 device and a conventional mercury sphygmomanometer.} The measured BPs for each person were approximately the same (each person was at the rest condition during BP measurement using both monitors, so approximate same values for BPs for each person were expected). \textcolor{black}{Specifically, the mean absolute error of SBP and DBP measurements using these two devices were 3.4 mmHg and 5.0 mmHg, respectively.}

\begin{figure}[!t]
	\centering
	\includegraphics[width=0.5\textwidth]{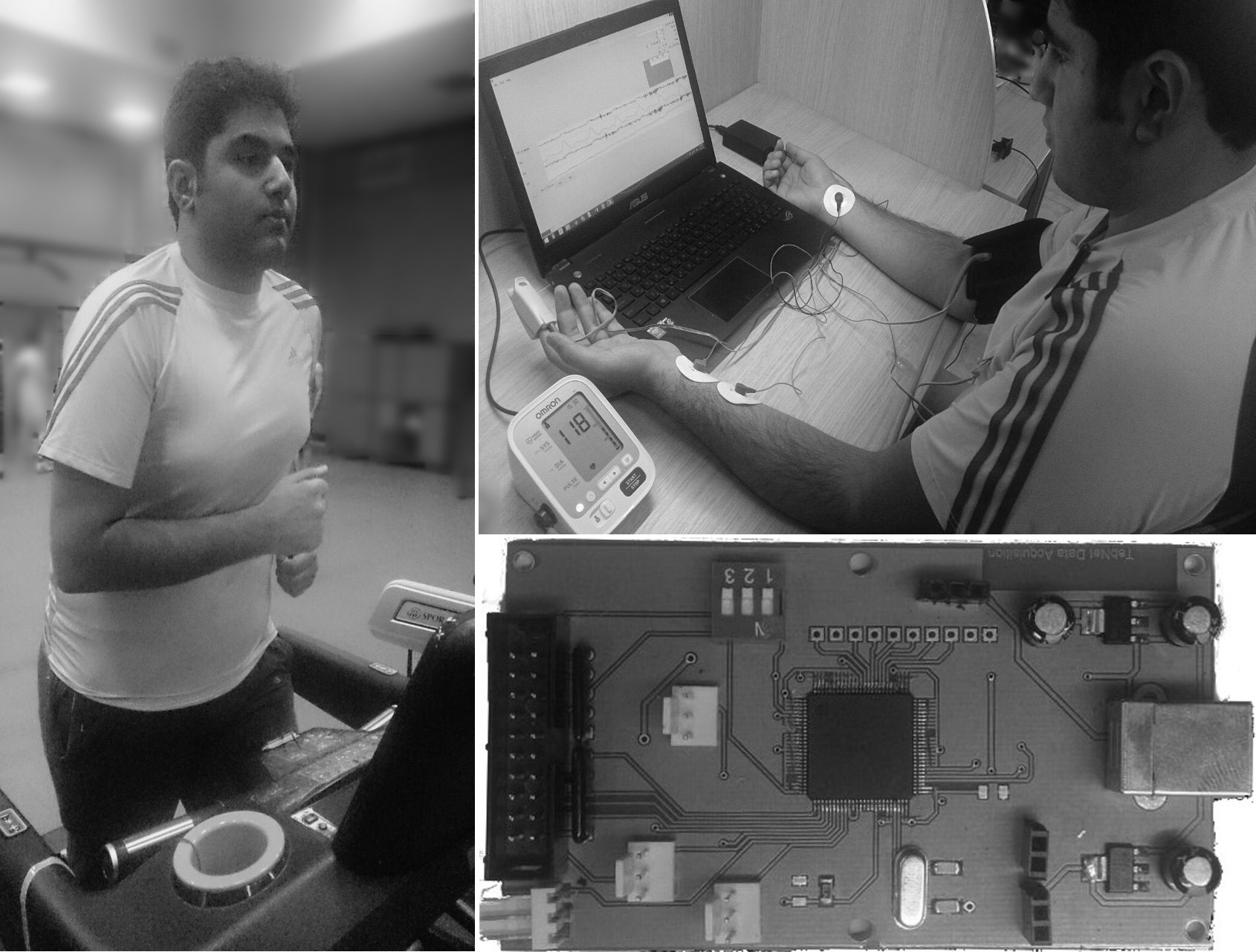}
	\caption{{\footnotesize An illustration of the data collection procedure and the data sampling board.}} \label{fig.real}
\end{figure}

\textcolor{black}{Since the PPG sensor was placed on the subject's the left hand, BP measurements were taken from the right hand to reduce the effects of BP measurement on the PPG signal (see Fig. \ref{fig.real}). The FSR was placed under the BP monitor cuff to measure the instantaneous cuff pressure. Using FSR is an effective and novel approach to measure the instantaneous cuff pressure synchronized with other vital signals. Further, it can be used to find the exact time instances of each BP measurement during the calibration process. Details of processing the FSR signal and using them will be discussed in Section \ref{section.PTT_PAT}.}

\subsection{Data Acquisition Procedure} \label{section.experimentalsetup}
\textcolor{black}{In total, we had 32 healthy subjects in the age range of 21-50 years (including 24 male and 8 female subjects).} Subjects were informed and agreed about the data collection procedure in advance. Also, the data collection procedure was approved by Health Center authorities at Sharif University of Technology. Among common interventions proposed in the literature for perturbing BP \cite{petrofsky1975aging}, \cite{al1997cardiovascular}, \cite{joseph2005slow}, \cite{parati1989comparison}, \cite{foo2006evaluation}, \cite{wong2009evaluation}, here, physical exercise is employed since it has been shown
to cause a sensible increase in both SBP and DBP up to
40 mmHg \cite{mukkamala2015toward}.
\textcolor{black}{For data collection for the calibration procedure, a same
supervised physical exercise, which was running about 3 minutes at the speed of 8 km/h, was used for all the subjects.}
\textcolor{black}{This physical exercise would increase their BP distinctly. When the physical exercise is finished, the subject would sit upright and the
data collection would begin}. During data collection, the subjects were told to remain stable with no movement, which was vital particularly for the FSR signal due to its sensitivity and also to
avoid motion artifacts for other signals.
They also were told not to speak because the consequent vibrations
in their chest could corrupt the PCG signal.
\textcolor{black}{For each subject, several reference BP measurements were performed. Since each subject sits during the data collection after the exercise,
the measured BPs are generally decreasing.} \textcolor[rgb]{0.00,0.00,0.00}{The data collection for the calibration purpose, including the
physical exercise, took around 15 minutes per subject. This calibration process is done only one time for each subject and after deriving parameters in the BP estimator model in \eqref{eq:P}, the BP can be estimated continuously.}\\

\begin{figure}[!t]
	\centering
	\includegraphics[width=0.4\textwidth]{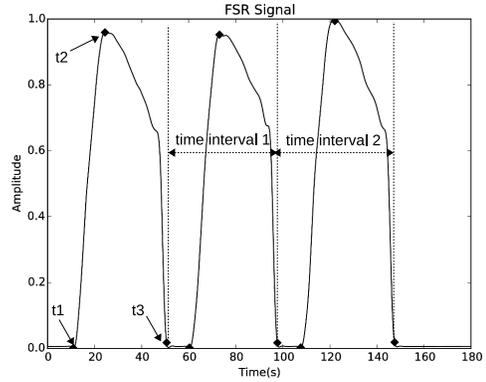}
	\caption{{\footnotesize The key moments on the FSR signal: $t_1$, $t_2$ and $t_3$. Time intervals corresponding to $t_3$ moments are \textcolor{black}{shown.}}} \label{fig.fsrdel}
\end{figure}

\section{Processing Pipeline} \label{section.processingpipeline}
A processing pipeline
is required for the estimation of
SBP and DBP. For this purpose, at first, a preprocessing stage
is devised in order to enhance the quality of the signals.
Afterwards, PTT/PAT values corresponding to each of the reference BP measurements are extracted. At last, using these values and the nonlinear model introduced in Section \ref{section.background}, SBP and DBP are estimated.

\textcolor{black}{“In this work, we have implemented the whole preprocessing, feature extraction, calibration, and estimation process using Python programming language. Also, the scikit-learn library \cite{pedregosa2011scikit} was used for training the regression models.”}

\subsection{Preprocessing on the Signals} \label{section.preprocess}
\textcolor{black}{For preprocessing on the signals, firstly, all four
signals are smoothed
using the median filter technique to reduce the impulsive noise effect.}
\textcolor[rgb]{0.00,0.00,0.00}{
As the median filtering is a non-linear technique,
the employed window width is limited to a few milliseconds (5 milliseconds).} \textcolor{black}{Next, all the signals are statically normalized. At the end,
frequency filtering is performed on the signals.}
\textcolor[rgb]{0.00,0.00,0.00}{
For frequency filtering, infinite impulse response (IIR) filters, which are more computational efficient than finite impulse response (FIR) filters, are used.
}
\textcolor[rgb]{0.00,0.00,0.00}{IIR filters, however, have non-linear phase responses, which may lead to distortions in signals
that adversely affect the accuracy of PTT/PAT measurements.}
To avoid
this pitfall, IIR filters are applied using a
non-causal forward-backward fashion (i.e. applying the filter in both directions), which produces a fixed zero phase response.

Particularly, for the FSR signal, because of
the relatively long time intervals between reference BP measurements, a low-pass filter is exploited. For the rest of the signals,
band-pass filtering is used.
\textcolor[rgb]{0.00,0.00,0.00}{
The filter specifications for each of the acquired signals in the data collection process are represented in Table \ref{table.filter}. Each filter's frequency response and order are set such that the main content of the corresponding input signal is retained.
}
\begin{table}
\centering
\caption{Filter Specifications}\label{table.filter}
\begin{tabular}{|c|c|c|c|c|}

     \hline
  Signal  & Filter  & Low   & High   & Order  \\
     Name &  Type &  Cutoff  &  Cutoff  &    \\
     \hline
     FSR Signal  & Bidirectional IIR,  & \textcolor[rgb]{0.00,0.00,0.00}{-} & \textcolor[rgb]{0.00,0.00,0.00}{0.3Hz} & 3\\
       & Low-Pass  &   &   & \\
     \hline
     PPG  & Bidirectional IIR, & 0.5    & 20  & 3\\
       & Band-Pass &   &   &\\
     \hline
     ECG  & Bidirectional IIR, & 1      & 40  & 3\\
       & Band-Pass &       &   &\\
     \hline
     PCG  & Bidirectional IIR, & 20     & 240 & 3\\
       & Band-Pass &       &   &\\
     \hline

\end{tabular}
\end{table}

\subsection{PTT/PAT Measurement and Post-Processing} \label{section.PTT_PAT}
\textcolor{black}{At the beginning, the three main signals (i.e. PCG, PPG and ECG) are divided to time intervals, where each interval corresponds to one of the BP measurements by the automatic BP monitor. There are three key moments in each reference BP measurement: \begin{enumerate}[i]\item When the cuff begins to inflate ($t_1$), \item When the cuff begins to deflate ($t_2$), \item When cuff deflation finishes and SBP and DBP values are measured and read by the automatic BP monitor ($t_3$). \end{enumerate} The FSR signal can distinguish these three moments since it represents instantaneous applied pressure by the cuff (see Fig. \ref{fig.fsrdel}). The $t_3$ moments are used to divide the main signals to the mentioned time intervals (see Fig. \ref{fig.fsrdel}).} Next, PTT and PAT values corresponding to each of reference BP measurements are calculated from corresponding time intervals.

\begin{figure}[!t]
\centering
\includegraphics[width=0.5\textwidth]{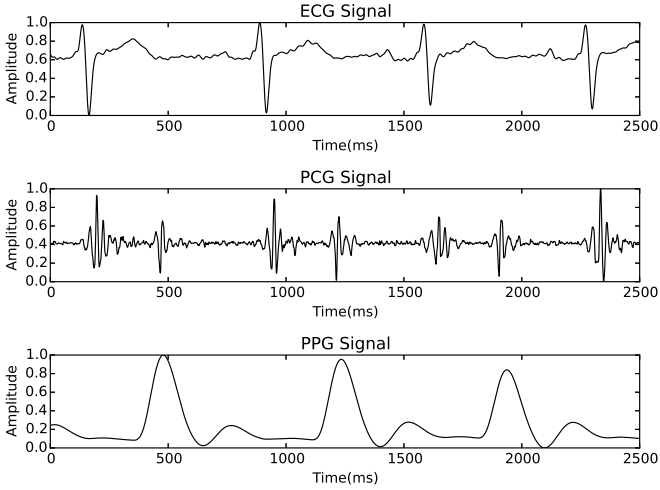}
\caption{{\footnotesize An example of a time window in which PAT/PTT calculation is done by analysis of ECG, PCG and PPG.}} \label{fig.window}
\end{figure}

Each time interval itself is divided into equal-sized time windows. The size of each time window is chosen such that at least 2 complete cardiac cycles of all the three main signals exist in each of them (e.g. 2.5 seconds). Moreover, a scaling is done in each time window to bring all the values into the range between 0 and 1 (see Fig. \ref{fig.window}).

\begin{figure}[!t]
\centering
\includegraphics[width=0.5\textwidth]{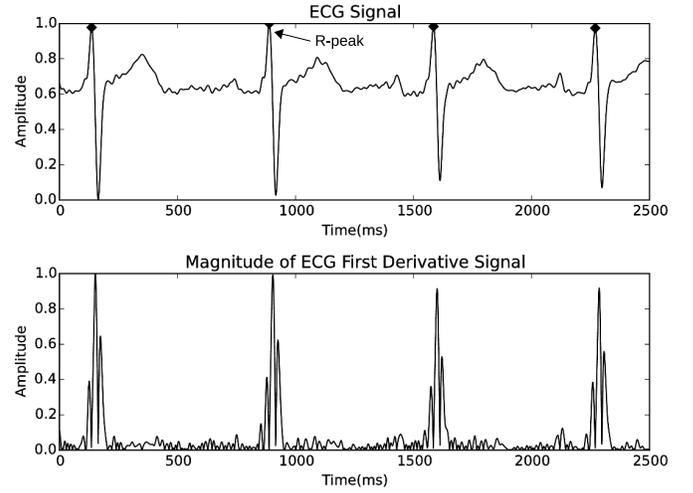}
\caption{{\footnotesize An example of ECG delineation. R-peak is found with the aid of the first derivative of ECG.}} \label{fig.ecgdel}
\end{figure}

\begin{figure}[!t]
\centering
\includegraphics[width=0.5\textwidth]{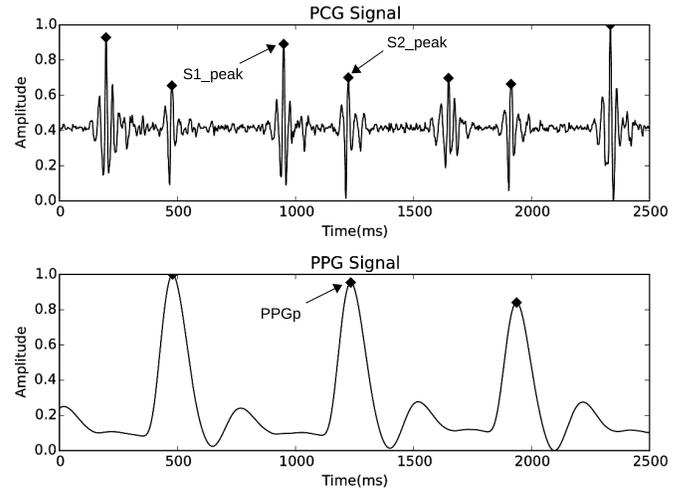}
\caption{{\footnotesize An example of PCG delineation. \mbox{S1-peak} is distinguished from \mbox{S2-peak} with the aid of the timing relation between PCG and PPG.}} \label{fig.pcgdel}
\end{figure}

For PTT/PAT measurement in each time window, the main signals should be delineated reliably. To determine the ECG R-peak and distinguish it from other peak types in ECG, the ECG itself and its first derivative are analyzed simultaneously. As the R-peak is more isolated than other peak types in ECG, the value of the first derivative of ECG is higher near the R-peak position (see Fig. \ref{fig.ecgdel}). \textcolor{black}{Accordingly, we applied a threshold of about 0.9 on the normalized ECG signal to find a few R-peak candidates and then we used the value of derivatives before and after each candidate to select the R-peak.}
For delineating PCG, however, the fact that S2-peak is closer than S1-peak to $PPG_p$ in each cardiac cycle can be employed to distinguish S1-peak from \textcolor[rgb]{0.00,0.00,0.00}{S2-peak} (see Fig. \ref{fig.pcgdel}). \textcolor{black}{Therefore, after finding a few candidate points for S1 and S2 peaks, we used their relative positions to the PPG systolic peak to decide on S1 and S2 assignments.}
$PPG_f$, $PPG_d$ and $PPG_p$ are determined through their definitions. Afterwards, PTT/PAT values corresponding to each of these distal points are measured. It is worth mentioning that, in each time window, if the proximal or
distal points cannot be delineated reliably, that specific window is excluded from our analysis.

\textcolor[rgb]{0.00,0.00,0.00}{
\textcolor{black}{At last, the average of the extracted PTT/PAT values from time windows in each time interval is used to give one PAT value and one PTT value for that time interval.} The obtained PTT average and PAT average in each time interval is corresponding to the reference BP measurement performed in that time interval. \textcolor[rgb]{0.00,0.00,0.00}{Doing this for all time intervals in data collection for each subject, we obtain reference BPs and associated PAT/PTT values for each subject. Using these pairs, the parameters in \eqref{eq:P} can be approximated for each individual.}
}

\subsection{Model Creation} \label{section.model}
As it was briefly mentioned in Section \ref{section.background}, in this paper, it is suggested to fit the non-linear BP-PTT relation of (\ref{eq:P}) to present a more accurate estimation of blood pressure values. Here, the well-known Gradient Descent (GD) optimization algorithm is employed to fit the nonlinear model of (\ref{eq:P}) using the collected calibration data. In specific, GD optimizes a multi-variable function $F(X)$, by starting from an initial guess about parameters ($X_{0}$), computing the gradients with respect to parameters, and updating them using the following equation:
\begin{equation}
X_{n+1} = X_{n} - \gamma * \Delta F(X_{n}), 
\end{equation}
where $\Delta F(X_{n})$ is the gradient of the optimization function at point $X_{n}$, and multiplied by $\gamma$, which is called the learning rate.

\textcolor{black}{Here, analytical derivatives of (\ref{eq:P}) are calculated with respect to the calibration model parameters. Afterwards, using these gradient values, the GD optimization algorithm is employed for the sake of recursively updating the model parameters. Apart from this, }
while in the GD algorithm initial values are usually selected randomly, here, due to the limited number of training BP and PTT pairs, the selection of the initial parameter values plays an important role in the fitting accuracy and the performance of the final model. \textcolor[rgb]{0.00,0.00,0.00}{Here, only for the initial parameter values, we assume $a_{1}$ is 0. Therefore, the PTT-BP relationship in \eqref{eq:P} can be simplified as:
\begin{equation}\label{eq:Pest}
BP = a_{0} + \sqrt{a_{2}} PTT^{-1} .
\end{equation}
Then, the initial values of $a_{0}$ and $a_{2}$ coefficients in \eqref{eq:P} are calculated using the least squares method in \eqref{eq:Pest}. Finally, starting with these initial points ($a_{1}$ = 0, $a_{0}$ and $a_{2}$ calculated using least squares method in \eqref{eq:Pest}), with an appropriate learning rate (e.g., 0.01), the gradient descent optimization method is used to determine $a_{0}$, $a_{1}$ and $a_{2}$ parameters for each individual, iteratively. This choice of initial values for parameters of \eqref{eq:P} have provided desirable results, as will be presented in Section \ref{section.results}.
}

\begin{table*}
  \centering

    \renewcommand{\arraystretch}{1.15}

  \caption{\textcolor{black}{The SBP and DBP Estimation Results Using The Proposed Nonlinear Estimator Model in \eqref{eq:P}}}\label{table.results}
  \begin{tabular}{c|c|c|c|c|c|c|c|c|}
  \cline{2-9}
  & \multicolumn{4}{c|}{The Performance of SBP Estimation} & \multicolumn{4}{c|}{The Performance of DBP Estimation}\\
  
  \hline
  \multicolumn{1}{|c|}{Exploited Index}&\textcolor{black}{ME}&MAE&STD&r&\textcolor{black}{ME}&MAE&STD&r\\
  
  \multicolumn{1}{|c|}{for Estimation}&\textcolor{black}{(mmHg)}&(mmHg)&(mmHg)&&\textcolor{black}{(mmHg)}&(mmHg)&(mmHg)&\\
  \hline
  \multicolumn{1}{|c|}{$PAT_f$}&\textcolor{black}{-0.21}&7.92&12.16&0.81&\textcolor{black}{0.73}&5.93&8.72&0.64\\

  \hline
  \multicolumn{1}{|c|}{$PAT_d$}&\textcolor{black}{0.80}&4.98&7.16&0.94&\textcolor{black}{1.59}&4.82&5.47&0.84\\

  \hline
  \multicolumn{1}{|c|}{$PAT_p$}&\textcolor{black}{0.12}&4.71&6.15&0.95&\textcolor{black}{1.31}&4.44&5.36&0.84\\

  \hline
  \multicolumn{1}{|c|}{$PTT_f$}&\textcolor{black}{0.69}&8.33&12.85&0.80&\textcolor{black}{1.37}&5.92&8.08&0.67\\

  \hline
  \multicolumn{1}{|c|}{$PTT_d$}&\textcolor{black}{0.22}&6.17&9.36&0.88&\textcolor{black}{1.09}&3.91&5.12&0.83\\

  \hline
  \multicolumn{1}{|c|}{$PTT_p$}&\textcolor{black}{-0.28}&6.22&9.44&0.89&\textcolor{black}{1.03}&3.97&5.15&0.84\\

  \hline
  \end{tabular}
\end{table*}

\section{Results} \label{section.results}
\textcolor[rgb]{0.00,0.00,0.00}{
For testing the proposed approach, per subject, each time, we leave one of the extracted pairs of reference BPs and associated PTT/PAT values aside, and train the nonlinear model of \eqref{eq:P} with the remaining pairs of that subject. Then, that subject's BP is estimated using the obtained estimator model and the PAT or PTT index of the left-out pair. Then, the estimated BP value is compared with the actual reference BP value of the left-out pair. There are 5 or 6 extracted (PAT/PTT , reference BP) pairs for each subject in the data collection process. Therefore, we have 5 or 6 BP estimations per subject. \textcolor{black}{In total, there are 173 BP estimations for all 32 subjects in the study.}
}
\textcolor[rgb]{0.00,0.00,0.00}{
\subsection{Comparison between different distal timing references on the PPG signal} \label{section.modelcom}
Table \ref{table.results} presents the performance of the proposed model for different variations of PAT/PTT values, corresponding to using different characteristic points on the PPG signal as the distal timing reference. \textcolor[rgb]{0.00,0.00,0.00}{Criteria for performance evaluation are \textcolor{black}{mean error (ME)}, mean absolute error (MAE) of estimation, standard deviation (STD) of estimation, and target-estimation correlation coefficient (r).} According to Table \ref{table.results}, some comparisons can be made. For instance, the estimation of the model, when $PPG_f$ is selected as the distal point \textcolor{black}{for measuring PTT or PAT}, is inferior to estimation using other characteristic points on PPG. Using both $PPG_d$ and $PPG_p$ for estimation, \textcolor{black}{performs nearly the same for both of SBP and DBP targets. \textcolor{black}{It is noteworthy to mention that the subject-specific parameters of (\ref{eq:P}) are determined for each individual separately using the suggested calibration procedure.}}
}



\begin{figure}
    \centering
    \begin{subfigure}[t]{0.50\textwidth}
        \centering
        \includegraphics[width = 1\textwidth]{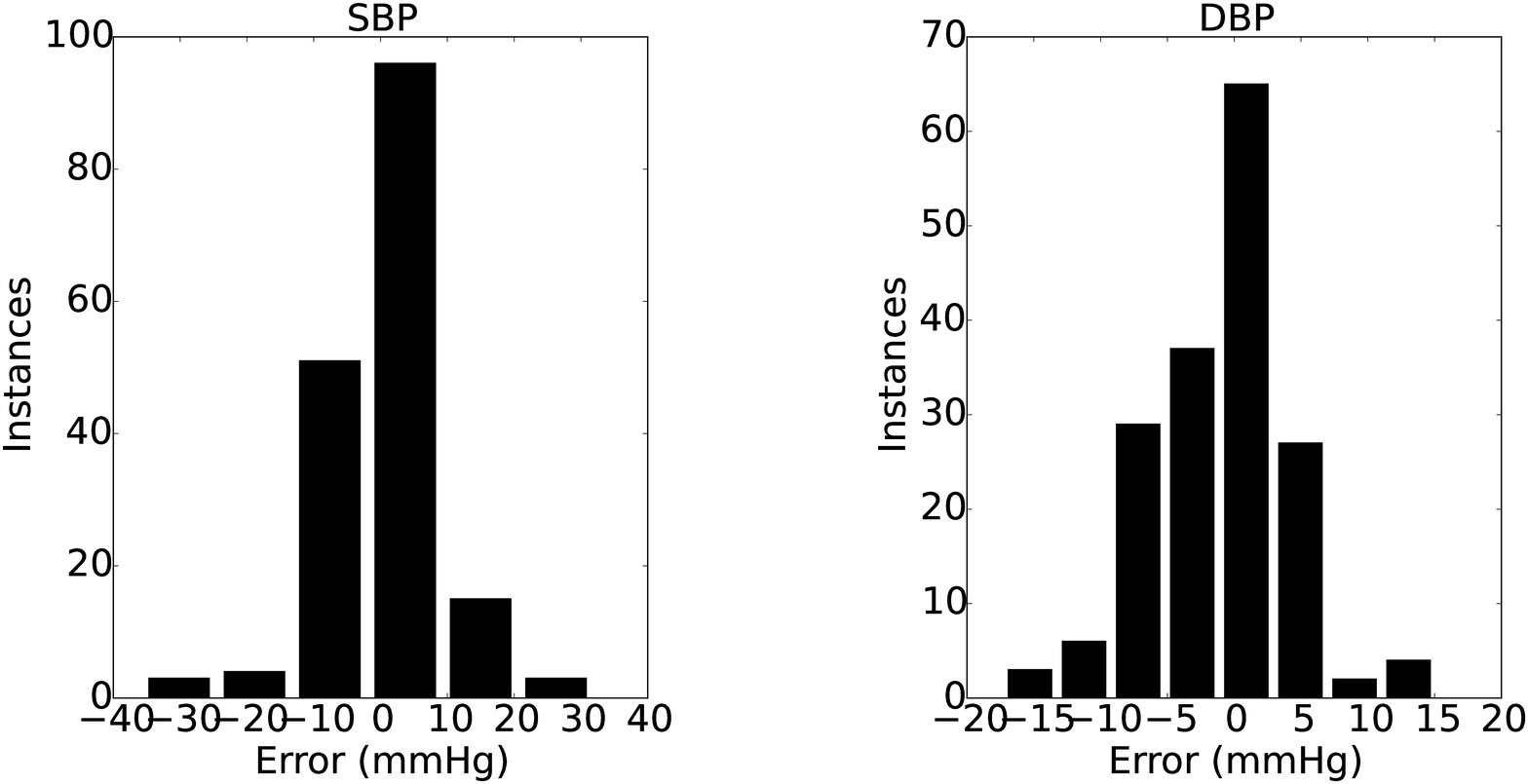}
        \caption{}
        \label{fig.his_PTT}
    \end{subfigure}
    \begin{subfigure}[t]{0.50\textwidth}
        \centering
        \includegraphics [width = 1\textwidth]{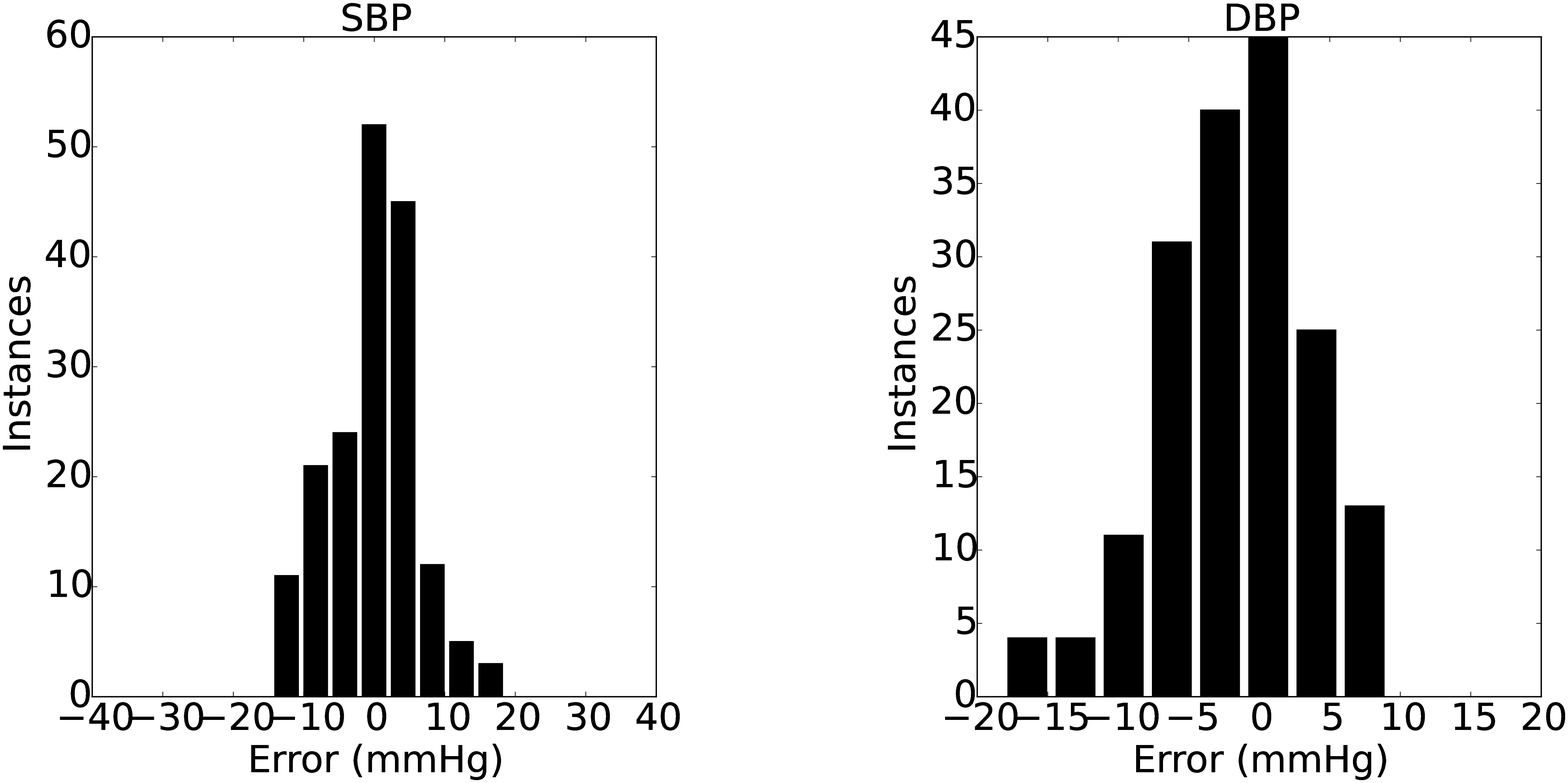}
        \caption{}
        \label{fig.his_PAT}
    \end{subfigure}
    \caption{\footnotesize{\textcolor{black}{Error histograms for SBP and DBP targets (a) using PTT index and (b) using PAT index.}}}\label{fig.his}
\end{figure}

\begin{figure}
    \centering
    \begin{subfigure}[t]{0.50\textwidth}
        \centering
        \includegraphics[width = 1\textwidth]{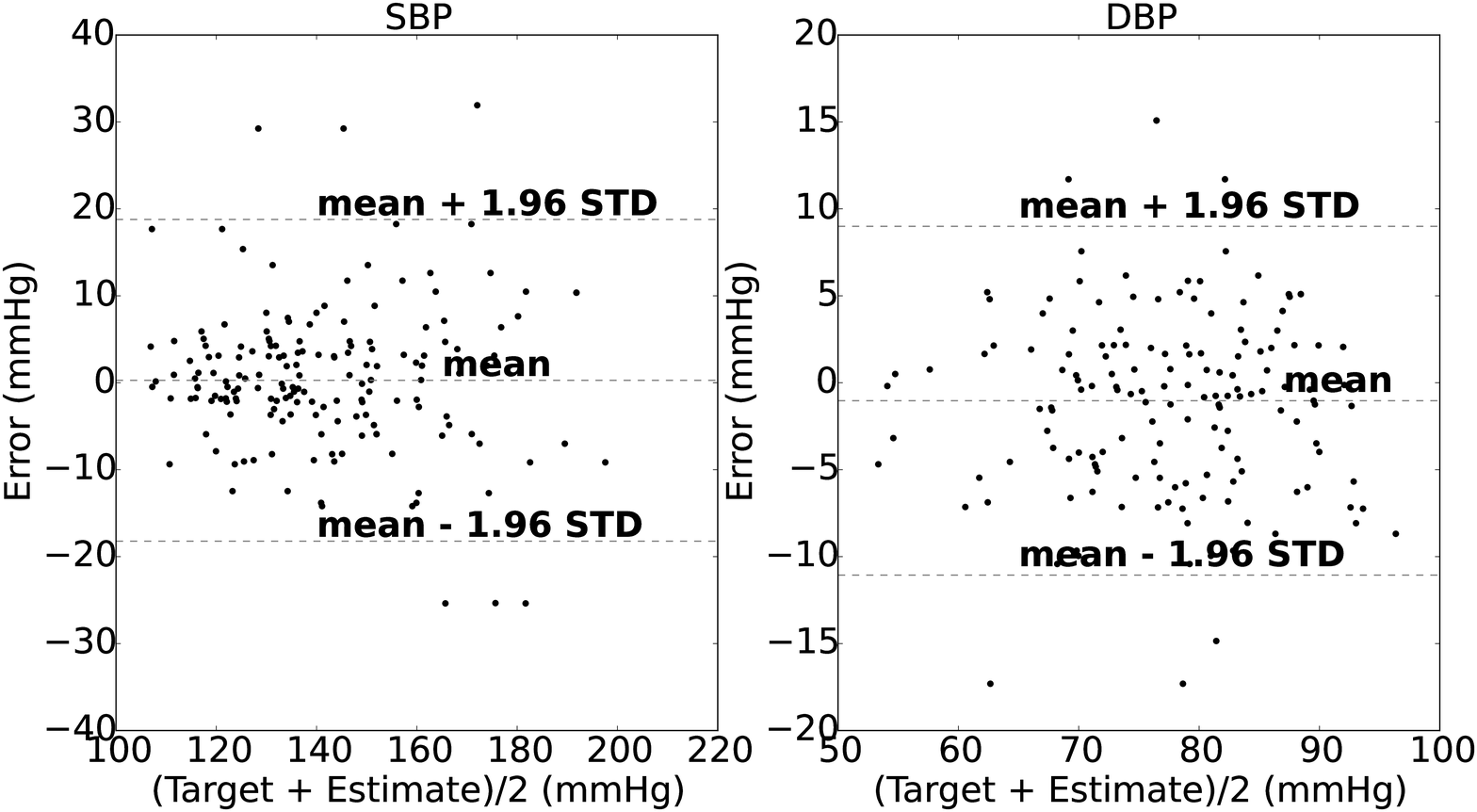}
        \caption{}
        \label{fig.bland_PTT}
    \end{subfigure}
    \begin{subfigure}[t]{0.50\textwidth}
        \centering
        \includegraphics [width = 1\textwidth]{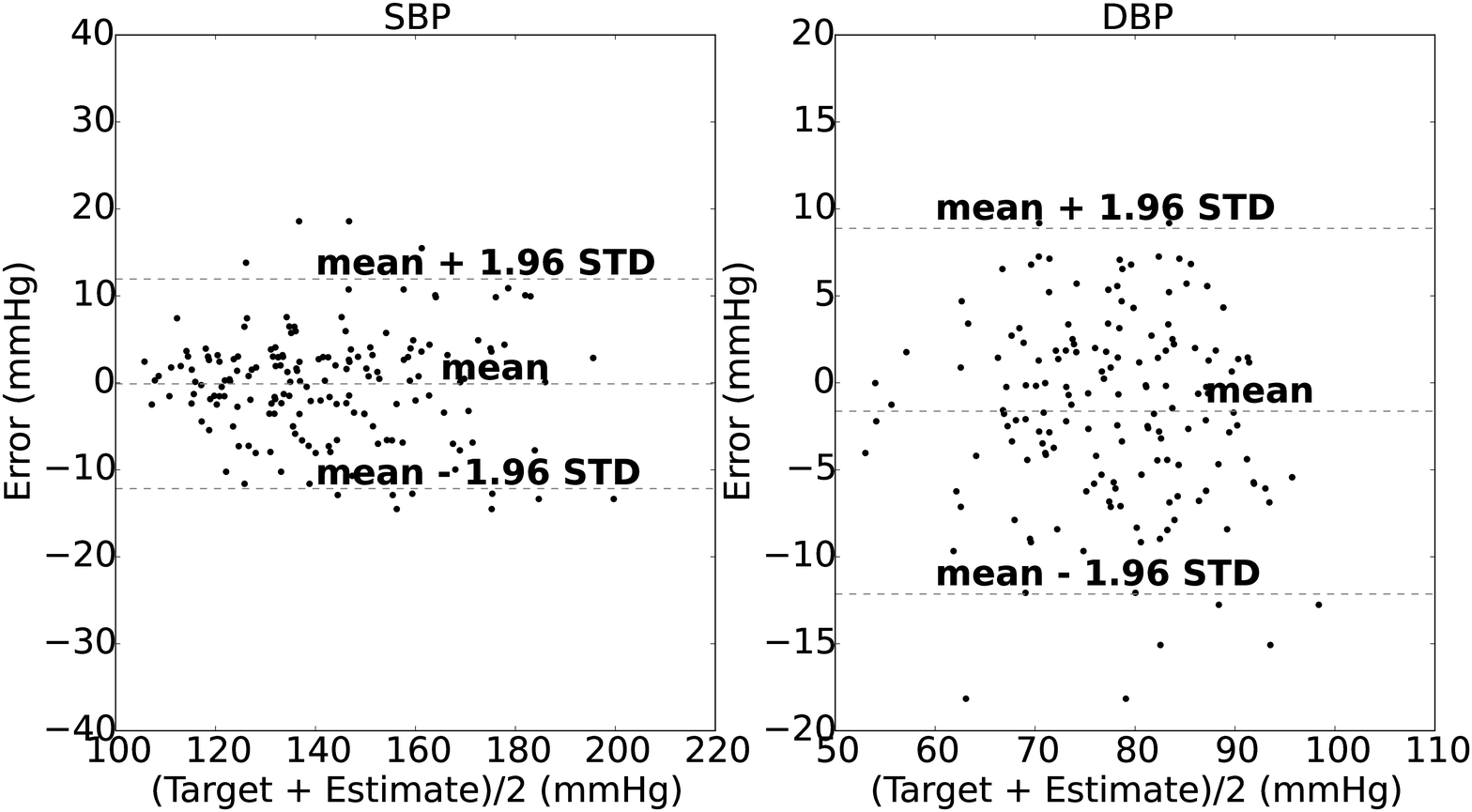}
        \caption{}
        \label{fig.bland_PAT}
    \end{subfigure}
    \caption{\footnotesize{\textcolor{black}{Bland-Altman plots for SBP and DBP targets (a) using PTT index and (b) using PAT index.}}}\label{fig.bland}
\end{figure}


\begin{figure}[!t]
    \centering
    \begin{subfigure}[t]{0.50\textwidth}
        \centering
        \includegraphics[width = 1\textwidth]{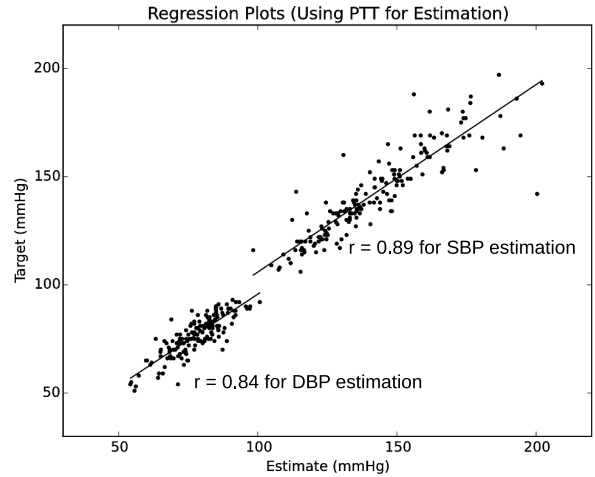}
        \caption{}
        \label{fig.reg_PTT}
    \end{subfigure}
    \begin{subfigure}[t]{0.50\textwidth}
        \centering
        \includegraphics [width = 1\textwidth]{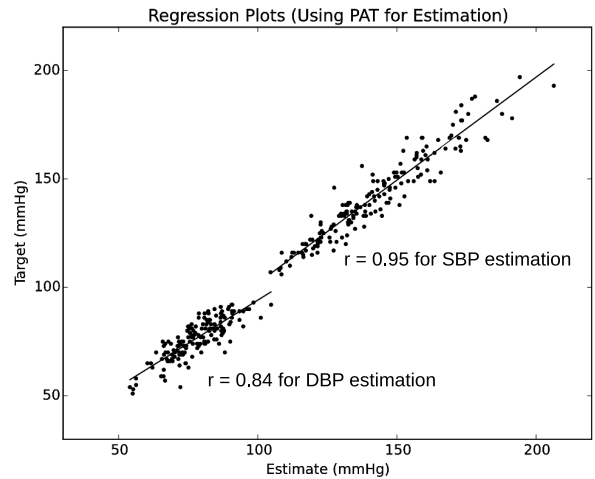}
        \caption{}
        \label{fig.reg_PAT}
    \end{subfigure}
    \caption{\footnotesize{\textcolor{black}{Regression plots for SBP and DBP targets using (a) PTT index and (b) PAT index.}}}\label{fig.reg}
\end{figure}

\begin{figure}
    \centering
    \begin{subfigure}[t]{0.50\textwidth}
        \centering
        \includegraphics[width = 1\textwidth]{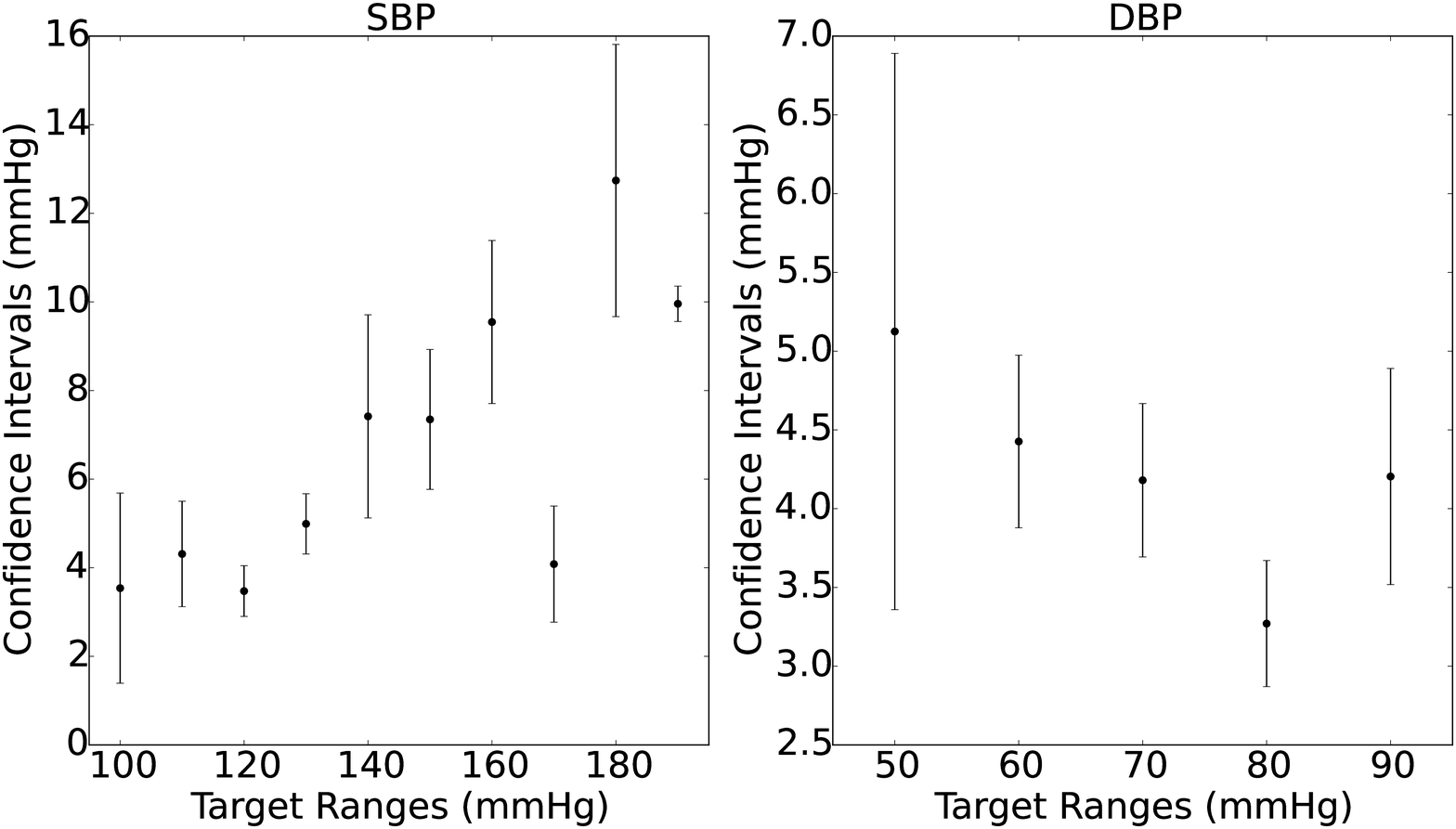}
        \caption{}
        \label{fig.CI_PTT}
    \end{subfigure}
    \begin{subfigure}[t]{0.50\textwidth}
        \centering
        \includegraphics [width = 1\textwidth]{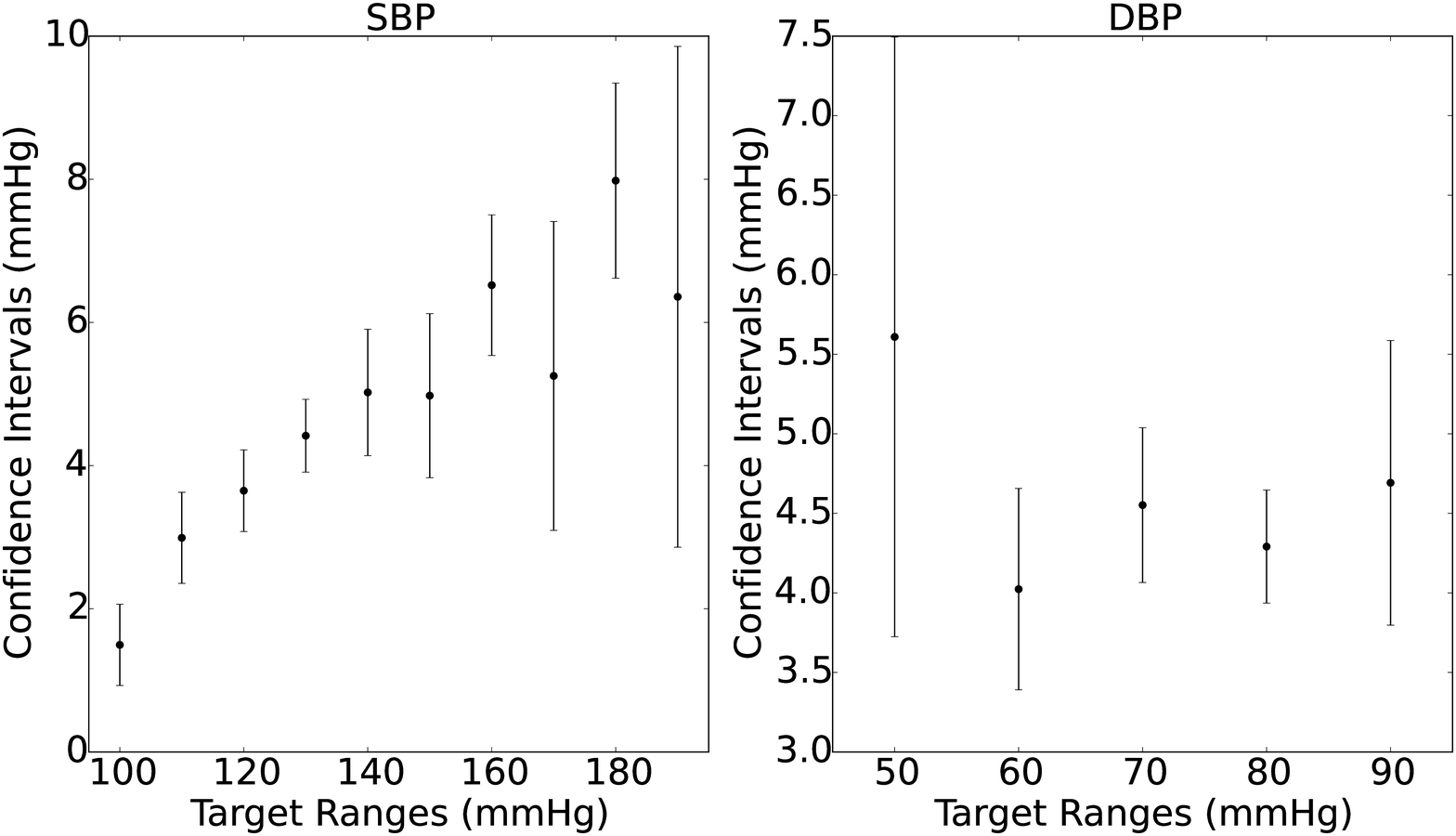}
        \caption{}
        \label{fig.CI_PAT}
    \end{subfigure}
    \caption{\footnotesize{\textcolor{black}{95\% confidence intervals for SBP and DBP targets (a) using PTT index and (b) using PAT index.}}}\label{fig.CI}
\end{figure}

\subsection{PAT Versus PTT} \label{section.patptt}
\textcolor[rgb]{0.00,0.00,0.00}{
For having a comparison between PAT and PTT in terms of their accuracy in both SBP and DBP estimations, \textcolor{black}{$PPG_d$ is used as the distal timing reference.} As it is evident, \textcolor{black}{for SBP estimation, PAT performs better than PTT, while the accuracy of DBP estimation using PTT is higher in compared to using PAT.}
}

\textcolor[rgb]{0.00,0.00,0.00}{Fig. \ref{fig.his} demonstrates the histograms of errors for SBP and DBP estimations. Here, as Table \ref{table.results} suggests, $PTT_p$ index or $PAT_p$ index are used for estimations. According to Fig. \ref{fig.his}, it can be seen that error values are distributed around zero. Based on these histograms, a major portion of the error values lie within the threshold of $\pm$ 10 mmHg for both of SBP and DBP targets.}

\textcolor{black}{Fig. \ref{fig.bland} \textcolor{black}{presents} the Bland-Altman plots for \textcolor{black}{SBP and DBP} estimations. \textcolor{black}{According to Fig. \ref{fig.bland}, a wide range of target BPs are able to be estimated with low errors (ranges for SBP and DBP targets are from 106 mmHg to 197 mmHg and from 51 mmHg to 93 mmHg, respectively.).} Also, the limits of agreement (mean $\pm$ 1.96$\times$STD) are indicated. \textcolor{black}{Using the PTT index, these limits are -0.28 $\pm$ 1.96$\times$9.44 mmHg for SBP and 1.03 $\pm$ 1.96$\times$5.15 mmHg for DBP. Using the PAT index, these limits are  0.12 $\pm$ 1.96$\times$6.15 mmHg for SBP and 1.31 $\pm$ 1.96$\times$5.36 mmHg for DBP.}}

\textcolor[rgb]{0.00,0.00,0.00}{The estimation versus target regression plots for BP estimations using PTT index or PAT index are \textcolor{black}{shown} in Fig. \ref{fig.reg}. \textcolor{black}{Using PTT index for estimation, r values for SBP and DBP estimations, \textcolor{black}{obtained from all the subjects}, are 0.89 and 0.84, respectively, where with the aid of PAT index, r value for SBP and DBP estimations are 0.95 and 0.84, respectively.} \textcolor{black}{Consequently, there is a close correlation between estimated and target BPs for both of SBP and DBP.} The strong correlation in the case of DBP is of great importance since generally, in literature, correlation coefficients of DBP estimations are \textcolor{black}{distinctly} less than those of SBP estimations (see Section \ref{section.comparison}).}

\textcolor{black}{Furthermore, as we are making some estimations and measurements in this study, and sources of uncertainty from both hardware and software sides could contribute to the variability of results, it is interesting to investigate and determine the confidence intervals. In other words, we need to know the numerical margin around each estimation (measurement) showing how much we are confident about the validity of our measured values compared to the actual  real values. Here, we divide the range of target values to 10 mmHg intervals, and in each of these intervals, we calculate 95\% confidence intervals (i.e., the margins we are 95\% confident that the difference between the real value and measured value lie within that). Fig. 11 shows these confidence intervals for both SBP and DBP targets. Also, since Fig.11 shows confidence intervals for wide ranges of SBP and DBP values, for any new patient with SBP and DBP values in these ranges, confidence of intervals of measurements are approximated. Based on the results in Fig. 11, considering the wide range of parameters involved in the BP estimation process, the achieved confidence interval are reasonable.}

\textcolor{black}{In addition, as a comparison between male and female subjects, MAEs for male and female subjects, for SBP estimation, are 6.50 mmHg versus 5.86 mmHg using PTT index, and 4.76 mmHg versus 4.64 mmHg using PAT index. Also, for DBP estimation, these values are 3.85 mmHg versus 4.00 mmHg using PTT index, and 4.34 mmHg versus 4.57 mmHg using PAT index. Therefore, no significant differences are observed between male and female subjects in terms of the accuracy of the proposed model.}

\begin{table*}

\renewcommand {\thefootnote} {\fnsymbol{footnote}}

\centering
\renewcommand{\arraystretch}{1.25}
\caption{\textcolor{black}{Comparison with Other Works} \protect\footnotemark[1]}
\label{table.otherwork}
\scalebox{0.85}{
	\begin{tabular}{|c|c|c|c|c|c|c|c|c|c|c|c|}
		\hline
		Work & Number of & \multicolumn{2}{c|}{Estimation Methodology} &
		\multicolumn{4}{c|}{SBP} & \multicolumn{4}{c|}{DBP}\\
		\cline{3-12}
		&Subjects&Signals (Index)&\textcolor{black}{Calibration} &\textcolor{black}{ME} &MAE&STD&r&\textcolor{black}{ME}&MAE&STD&r\\
		&&&\textcolor{black}{Interventions}&\textcolor{black}{(mmHg)}&(mmHg)&(mmHg)&&\textcolor{black}{(mmHg)}&(mmHg)&(mmHg)&\\
		\hline
		This Work &32&PCG \& PPG (PTT)&\textcolor{black}{Running Exercise}&\textcolor{black}{-0.28}&6.22&9.44&0.89&\textcolor{black}{1.03}&3.97&5.15&0.84\\
		\hline
		This Work &32&ECG \& PPG (PAT)&\textcolor{black}{Running Exercise}&\textcolor{black}{0.12}&4.71&6.15&0.95&\textcolor{black}{1.31}&4.44&5.36&0.84\\
		\Xhline{3\arrayrulewidth}
		\cite{wong2009evaluation}&14&ECG \& PPG (PAT)&\textcolor{black}{Treadmill Exercise}&\textcolor{black}{0.00}&7\footnotemark[2]&5.30&0.87&\textcolor{black}{0.00}&5\footnotemark[2]&2.90&0.30\\
		\hline
		\cite{gesche2012continuous}&63&ECG \& PPG (PAT)&\textcolor{black}{Cycle Ergometer Exercise}& -0.02\footnotemark[2]&8\footnotemark[2]&10.10&0.83&\multicolumn{4}{c|}{\textcolor{black}{DBP estimation not addressed in this paper.}}\\
		\hline
		\cite{kim2015ballistocardiogram}&15&BCG \& PPG (PTT)&\textcolor{black}{Deep Breathing +}&-0.03\footnotemark[2]&9\footnotemark[2]&8.58\footnotemark[2]&0.70& -0.02\footnotemark[2]&7\footnotemark[2]&5.81\footnotemark[2]&0.66\\
		&&&\textcolor{black}{Sustained Handgrip}&&&&&&&&\\
		\hline
		\textcolor{black}{\cite{kachuee2016cuff}}&\textcolor{black}{57}&\textcolor{black}{ECG \& PPG (PAT)}&\textcolor{black}{No Intervention}&-0.04\footnotemark[2]&\textcolor{black}{8.21}&\textcolor{black}{5.45}&\textcolor{black}{0.54}&-0.03\footnotemark[2]&\textcolor{black}{4.31}&\textcolor{black}{3.52}&\textcolor{black}{0.57}\\
		\hline
		\textcolor{black}{\cite{martin2016weighing}}&\textcolor{black}{22}&\textcolor{black}{BCG \& PPG (PTT)}&\textcolor{black}{Three Hemodynamic Interventions}&-0.4\footnotemark[2]&\textcolor{black}{11.8\footnotemark[3]}&\textcolor{black}{1.6}&\textcolor{black}{0.80}&-0.2\footnotemark[2]&\textcolor{black}{7.6\footnotemark[3]}&\textcolor{black}{0.5}&\textcolor{black}{0.80}\\
		\hline
		\textcolor{black}{modified \cite{gesche2012continuous}
		\footnotemark[4]}&32&\textcolor{black}{ECG \& PPG (PAT)}&\textcolor{black}{Running Exercise}&\textcolor{black}{0.85}&\textcolor{black}{14.02}&\textcolor{black}{11.11}&\textcolor{black}{0.78}&\textcolor{black}{1.31}&\textcolor{black}{4.99}&\textcolor{black}{4.77}&\textcolor{black}{0.82}\\
		\hline
	\end{tabular}
}
\begin{flushleft}
    \textcolor{black}{\hspace{0.15cm} \footnotemark[1]{\scriptsize The values represented with "-" are not either reported in corresponding other works or able to be estimated based on reported results.} \\}
	\hspace{0.15cm} \footnotemark[2] {\scriptsize Approximated from the corresponding Bland-Altman plots. \\}
	\textcolor{black}{\hspace{0.15cm} \footnotemark[3] {\scriptsize In \cite{martin2016weighing}, instead of MAE, root-mean-squared-error (RMSE)  is reported.} \\}
	\textcolor{black}{\hspace{0.15cm} \footnotemark[4] {\scriptsize Applying the BP estimator model in \cite{gesche2012continuous} to our own dataset to for a fair comparison. See the details in Section \ref{section.comparison}.} \\}
\end{flushleft}	
\end{table*}

\subsection{Comparison with Other Works} \label{section.comparison}
\textcolor[rgb]{0.00,0.00,0.00}{
\textcolor{black}{A comprehensive comparison between the proposed method in this paper and some of other works in literature, which use PTT or PAT index for BP estimation, is presented in Table \ref{table.otherwork}}. For each one, number of subjects as well as estimation methodology are mentioned. According to this table, in comparison with other methods using PAT index for BP estimation, the results for the PAT-based estimations in the proposed method in this paper are more \textcolor{black}{encouraging} for both of SBP and DBP estimations \textcolor{black}{(by comparing r values)}. The same conclusion can be made by comparing the PTT-based estimations in this work and other methods using PTT index to estimate BP. It is worth mentioning that DBP estimations in this \textcolor{black}{study}, using PTT or PAT, have \textcolor{black}{promising results} \textcolor{black}{in comparison to other studies}. This fact is valuable since in many studies in literature, compared to SBP, correlation coefficients associated with DBP are \textcolor{black}{considerably} less (see \cite{wong2009evaluation,hennig2013continuous,wong2009acute} for example).
} 

\textcolor{black}{Furthermore, since using different subjects and datasets can lead to different results, for a fair comparison, we exploited the BP estimator model in \cite{gesche2012continuous} and applied it to our own dataset. The BP estimator in \cite{gesche2012continuous} uses a general exponential function based on PWV, which is derived using PAT. The parameters of this general function are obtained by least squares fitting the function to a train group. In this paper, we used first \textcolor{black}{17 subjects} for this purpose. Then, the derived function is used to estimate the BP of a test group. Here, we used the remained \textcolor{black}{15 subjects} for this purpose. Also, \cite{gesche2012continuous} suggests to use a one-point calibration method to compensate the subject-specific errors of the general function. Here, as we have 5 or 6 pairs of PAT values and reference BP measurements per subject, we used the first pair for the calibration and the rest for the evaluation of the model. The results for both SBP and DBP estimations are shown in Table \ref{table.otherwork} (In \cite{gesche2012continuous}, only results for SBP estimations are reported). According to the results, by comparing the r values, the estimation accuracy of the proposed model in the paper is still higher, even after using our own dataset for the proposed model in \cite{gesche2012continuous}.}


\textcolor{black}{
\section{\textcolor{black}{Discussions and Limitations}} \label{section.discussion}
It is worth to mention the goal of using physical exercise in this paper was to obtain a number of reference BPs and PAT/PTT values with a significant variance to derive the parameters of the model in (\ref{eq:P}) more accurately because of reasonably different calibration points, provided by the  physical exercise. However, the physical exercise cause PEP and PTT to vary in the same directions \cite{martin2016weighing}, and therefore, since PTT is directly correlated to BP, the PAT can track BP more accurately compared to when we use other interventions such as cold pressor. Consequently, any approach that employs the same proposed model as in this paper but with other interventions for perturbing BP that change PEP and PTT in opposite directions, can affect the PAT-based results adversely.}

\textcolor{black}{
Furthermore, in this paper, we measured PTT/PAT values through the brachial artery. The stiffness of the brachial artery, and therefore PTT/PAT values, may vary through the changes in vasomotor tone (controlled by the sympathetic nervous system), and not necessarily changes in BP \cite{ding2016continuous}. Some studies have examined the effect of vasomotor tone on BP-PTT relationship \cite{liu2014attenuation}, and central PPG instead of peripheral PPG was suggested to reduce this effect \cite{sola2013noninvasive}. Some studies have done the measurement of PTT through larger, more elastic arteries including the aorta using a system in a way similar to a bathroom weighing scale \cite{martin2016weighing}. Nonetheless, most studies have not attempted to take this factor into account in the PTT-BP estimation model to improve the accuracy.
}

\textcolor{black}{One of the limitations of this work, which can be examined in future studies, is the fact that subjects used in this study were generally young and healthy subjects, while using a homogeneous dataset like this can affect the results. For instance, patients with cardiovascular diseases can have highly different heart rates over heartbeats, to the extent that the extraction of valid PTT and PAT values from vital signals could not be feasible. Therefore, one of the future directions could be investigating the proposed BP estimator model in this paper on more heterogeneous subjects having some specific cardiovascular diseases or wider age ranges.
}

\section{Conclusion} \label{section.conclusion}
\textcolor[rgb]{0.00,0.00,0.00}{
\textcolor{black}{In this paper, we have addressed the problem of continuous and cuff-less BP estimation.} \textcolor[rgb]{0.00,0.00,0.00}{PCG, ECG and PPG signals were acquired during the suggested data collection process, to obtain PTT and PAT indexes, which are utilized for BP estimation.} Each subject underwent a supervised physical exercise for perturbing BP in a sensible way for the calibration procedure. Exploiting the FSR signal was proposed for the sake of a precise synchronization between the reference BP measurements and the vital signals, which improved the calibration procedure significantly. The devised processing pipeline was employed for creating a BP estimator nonlinear model using PTT or PAT index for each individual. \textcolor{black}{The parameters of this BP estimator model are individual-specific and determined for each subject separately through the suggested calibration procedure.} \textcolor{black}{Comparing different distal timing references for measuring PTT or PAT, it can be inferred that using $PPG_d$ or $PPG_p$ could provide better correlations with both SBP and DBP.} \textcolor{black}{Using the PAT index for estimation, which is more widely used in literature, leads to  more precise SBP estimations, while the use of PTT performed better compared to using PAT for DBP estimations.} In summary, this study showed that the proposed method and the nonlinear estimator model, using either PTT index (obtained from PCG \& PPG signals) or PAT index (obtained from ECG \& PPG signals), can be employed for accurate estimation of both of SBP and DBP targets.
}


\end{document}